\documentclass[aps,pra,twocolumn]{revtex4}
\usepackage{longtable}
\usepackage{amssymb}
\usepackage{amsfonts}
\usepackage[pdftex]{hyperref}
\usepackage{hypcap} 
\usepackage{graphicx}
\usepackage{amsthm}
\usepackage{amsmath}
\usepackage{bbm}
\usepackage{amsbsy}
\usepackage{mathrsfs}

\def\refeqn#1{eq.\ (\ref{Equation::#1})}

\begin{document}
    \title{Bayesian estimation for selective trace gas detection}
    \author{John K. Stockton}
    \email{john.stockton@gmail.com}
    \author{Ari K. Tuchman}
    \affiliation{Entanglement Technologies, 3723 Haven Avenue, Menlo Park, California 94025}

\date{\today}

\begin{abstract}
We present a Bayesian estimation analysis for a particular trace gas detection technique with species separation provided by differential diffusion.  The proposed method collects a sample containing multiple gas species into a common volume, and then allows it to diffuse across a linear array of optical absorption detectors, using, for example, high-finesse Fabry-Perot cavities.  The estimation procedure assumes that all gas parameters (e.g. diffusion constants, optical cross sections) are known except for the number population of each species, which are determined from the time-of-flight absorption profiles in each detector.
\end{abstract}

\maketitle

\begin{figure}
\capstart
\includegraphics[width=3.0in]{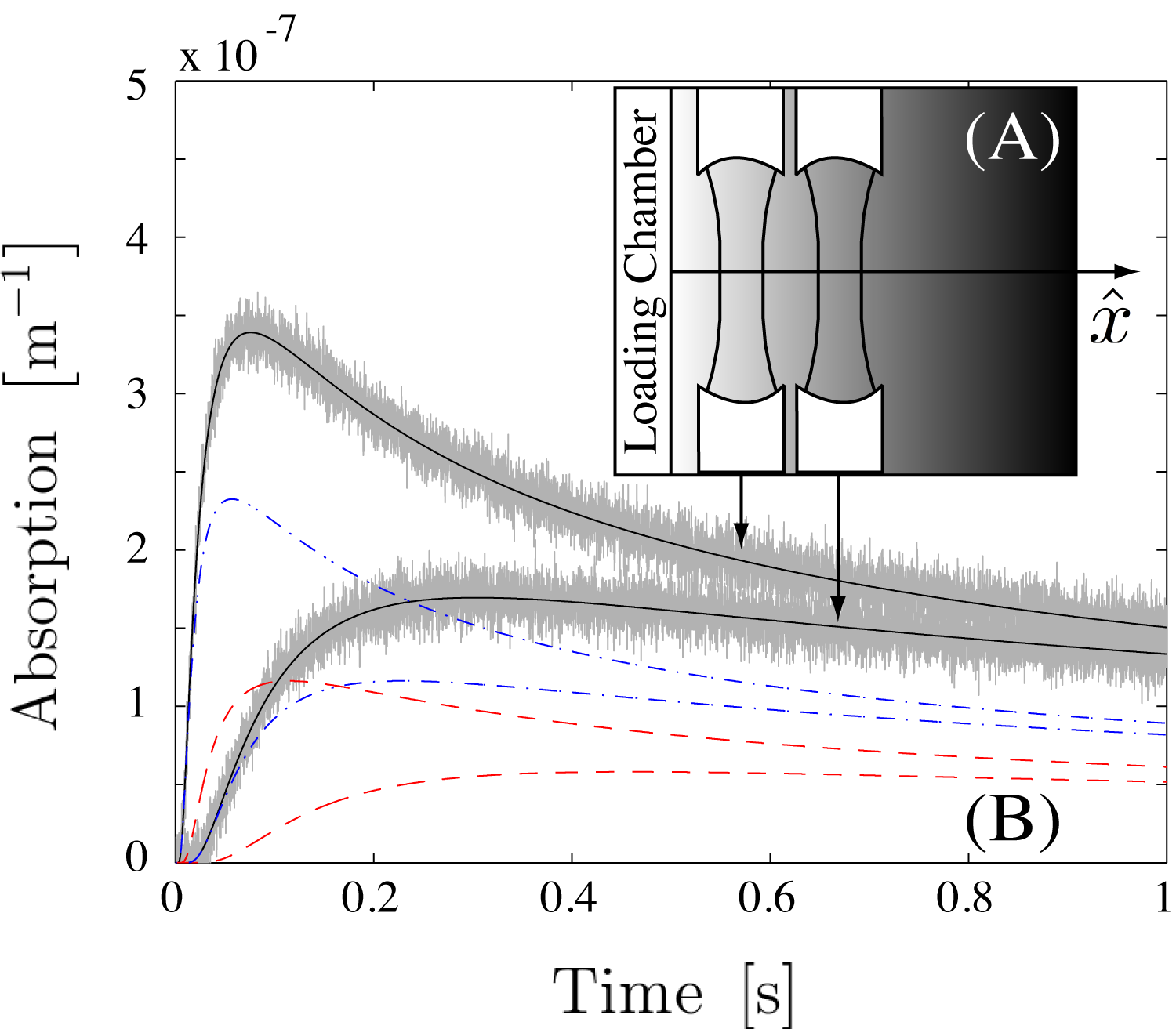}
\caption{(A) A gas mixture is released from a loading chamber and diffuses across two cavity modes, each of which generates an absorption signal. (B) The noisy plots show the raw absorption output from the two cavity measurements.  The dashed (red) curves denote the contribution from gas-1 and the dashed-dotted (blue) curves denote the contribution from gas-2.  The set of curves which peak first are for cavity-1 and the set which peak second are for cavity-2.  The chamber has a square uniformly filled transverse cross section $1$ cm per side.  Each cavity mode has a length $8 \ \textrm{mm}$, and a transverse width (along $x$) of $0.6 \ \textrm{mm}$.  The cavity modes are located along $\hat{x}$ at a distance of $1$ and $2$ mm from the origin.  The gas is collected into a collection volume of size $10\ \textrm{mm}^3$ ($0.1 \ \textrm{mm}$ width along x), where the initial concentrations are set at $1$ and $2$ ppb (parts per billion).  The optical absorption cross-section is the same for each gas and set to $\tilde{\sigma}_i=10^{-18}\ \textrm{cm}^2$ which is a typical value for the absorption of infrared radiation by molecules such as ethane $(C_2H_6)$. The diffusion constant for gas-1 is $D_1=0.044\ \textrm{cm}^2/\textrm{s}$ (the approximate value for ethane at room temperature and atmospheric pressure) and gas-2 is set a factor of two larger.  The white measurement noise corresponds to a $\sigma_M=10^{-12}\ \textrm{cm}^{-1}/\sqrt{\textrm{Hz}}$ absorption sensitivity.
} \label{Figure::1}
\end{figure}

Trace gas detection techniques have wide-ranging applications including explosives detection \cite{Todd2002} and medical diagnostics \cite{Breathalyzer2006}.  Individual sensors within a given system should have adequate sensitivities for measuring physical properties of the gas, but must also demonstrate the selectivity necessary to distinguish between the component gas species.   For example, if the optical absorption by a uniformly mixed sample of two gas species is measured at a single optical frequency, one can only determine the (scaled) sum of the two populations, but not the individual populations.  There are several approaches for obtaining selectivity, including the use of multiple probe frequencies along with a knowledge of the different gas spectra.  This can be done with sequential measurements with a tunable laser \cite{Todd2002} or simultaneously, as in recent work with femtosecond lasers \cite{Breathalyzer2006}.  Alternatively, one can spatially separate the species as in, for example, the industrial workhorse methods of mass spectrometry and gas chromatography.

Given a method that is both sensitive and selective, the task remains to estimate the populations of each gas within the original sample from the raw measurement data.  Bayesian estimation provides the most natural framework to update one's state of knowledge according to the acquired data.  In this approach, we start with a prior number distribution for each species, then iteratively update the entire joint probability distribution, using the measurement results and Bayes' rule \cite{Jacobs1996}.  The widths of this joint distribution will narrow in time due to this procedure, reflecting the acquisition of knowledge. As long as the assumptions going into the experimental model are accurate, the Bayesian approach will be optimal, in contrast to more common heuristic approaches.  Bayesian methodology is becoming increasingly relevant in many areas of physics and technology \cite{McHale2004, Stockton2007} as experiments reach such levels of sophistication that crude estimation techniques severely limit performance.  For example, in single-particle tracking experiments, Bayesian estimation can be used to extract the particle type based on the estimation of its diffusion constant \cite{McHale2004}.

In this paper, we propose a physical device capable of selectively measuring individual trace gas concentrations from a mixture containing multiple gas species, based on characteristic diffusion times.  We also present the Bayesian estimation formalism for mapping the device's raw measurement output to the individual trace gas populations.  The essential configuration of the diffusion spectrometer we propose is displayed in fig. 1A.  First, a sample containing a number of gas species is collected within the initial common volume and then released to diffuse into the sensor chamber, which is filled with a buffer gas and lined with a linear array of detectors.  We consider explicitly a series of high-finesse optical cavities which measure optical absorption at a single frequency.  However, the Bayesian method presented here can be easily extended to different trace gas detectors beyond the optical cavity, with varying forms of space- or frequency-based selectivity. The difference in diffusion constants causes the different species of atoms to transiently fill the chamber at differing rates, allowing for a selectivity not possible in a configuration where all of the species are mixed homogenously.  Fig. 1B displays the measurement output for each cavity along with the contributions from each gas at each point in time.  Note that even a device with only one cavity displays selectivity due to differential diffusion, however increasing the cavity number is beneficial both for the extra information, and for allowing common-mode cancellation of technical cavity noise (which is a common motivation for gradiometer configurations).

More specifically, an initial volume of gas containing $N_i$ particles of gas $i$ is released into the chamber.  We assume each gas has a distinct diffusion constant $D_i$ (in units of m$^2/$sec) and optical scattering cross section $\tilde{\sigma}_i$ (in units of m$^2$), and that these quantities are known for each potential gas in our library.  We assume that the gases uniformly fill the transverse direction, such that the distribution is only given along the direction of propagation as
\begin{eqnarray}
p_{i}(x,t)=2\frac{\exp\left(-x^2/2\sigma_i^2(t)\right)}{\sqrt{2\pi\sigma_i^2(t)}}.
\end{eqnarray}
The number density at $x$ is given by $\rho_i(x,t)=N_i p_{i}(x,t)/A$ where $A$ is the cross-sectional area of the tube.  The optical absorption coefficient is defined as $\alpha_i(x,t)=\rho_i(x,t)\tilde{\sigma}_i$ (in units of $1/$m).  The width of the distribution at time $t$ is given by $\sigma_i^2(t)=\sigma_i^2(0)+2 D_i t$.  We place a number of sensors in a linear array along $\hat{x}$, and the measurement output for a sensor $j$ at location $x_j$ is given by
\begin{eqnarray}
y_{j}(t)&=&\sum_i  C_{j,i}(t)N_i+\sigma_M \xi_{j}(t)
\end{eqnarray}
where the absorption at $x_j$ due to gas $i$ is $\alpha_{i}(x_j,t)=C_{j,i}(t)N_i$, 
$C_{j,i}(t)=p_i(x_j,t)\tilde{\sigma}_i/A$, and $\xi_j(t)$ is a white noise term (of variance $1/dt$).  We assume that each detector has the same absorption sensitivity $\sigma_M$, which for high-finesse optical Fabry-Perot cavities can approach $10^{-14} \ \textrm{cm}^{-1}/\sqrt{\textrm{Hz}}$ \cite{Ye1998}. 

In full generality, the Bayesian approach would use each successive measurement result, the assumed model, and Bayes' rule to update the collective distribution that describes our state of knowledge about the gas number populations.  This description can be simplified if we assume that the distribution begins and remains Gaussian.  In this case, we do not have to evolve the entire distribution, but merely a small set of moments, or more specifically, the number estimates $\vec{N}_e$ (the distribution means), and the covariance matrix, $\Sigma(t)$ (the distribution widths) \cite{Jacobs1996}.  The diagonal elements of the covariance matrix self-consistently represent our uncertainty in the populations and the average error of our estimates, $\langle (N_{e,i}-N_i)^2\rangle$, which the Bayesian estimator serves to minimize.

Technically, this Gaussian approach is not always ideal.  First, the Gaussian tail allows for the possibility of a non-physical negative population.  Second, for many trace gas scenarios the distribution is naturally bimodal with one peak at zero (no gas present) and another peak at a non-zero value (at the typical concentration level when the gas is present). Nevertheless, the Gaussian approximation is illustrative and simplifies the numerical treatment.  To indicate that a certain level of gas is `detectable' in this simplified framework, we require that the width of the distribution be less than the typical concentration value when the gas is present.

The covariance matrix is initialized at $\Sigma(0)=\Sigma_0$ which is a diagonal matrix with elements corresponding to the prior level of number uncertainty for each species.  According to the Bayesian procedure with the Gaussian assumption, the equation of motion for the covariance matrix, or Riccati equation, is the deterministic nonlinear matrix differential equation
\begin{eqnarray}
\frac{d\Sigma(t)}{dt}=-\frac{\Sigma(t) C^{T}(t) C(t) \Sigma(t)}{\sigma_M^2}\label{Equation::Riccati}.
\end{eqnarray}
This solution of this equation can be written \cite{Reid1972}
\begin{eqnarray}
\Sigma(t)&=&\left(\Sigma(0)^{-1}+\int_0^t \frac{C^{T}(s) C(s)}{\sigma_M^2} ds \right)^{-1}\label{Equation::RiccatiSolution}.
\end{eqnarray}
When the initial spatial gas distributions are approximated as delta functions ($\sigma_i^2(0)=0$), the integral can be represented in terms of the exponential-integral function. 

This treatment results in the following estimator which maps the noisy measurement results to the best-guess estimates for the population of each species in the sample volume:
\begin{eqnarray}
\frac{d \vec{N}_{e}(t)}{dt}&=&K_O(t) \left(\vec{y}(t)-C(t) \vec{N}_{e}(t)\right).
\end{eqnarray}
Here $\vec{y}(t)-C(t) \vec{N}_{e}(t)$ represents the `innovation' or the extent to which the measurement results deviate from our expectation. The matrix $K_O(t)=\Sigma(t) C^{T}(t)/\sigma_M^2$ represents the `observer gain' which weights the importance of the new measurement results \cite{Jacobs1996}.  Note that the covariance matrix not only reflects the expected performance but also tells us how to tailor this observer gain in time. Namely, the more information that is acquired, the less relative importance we give to the latest data point.

\begin{figure*}
\capstart
\includegraphics[width=6.0in]{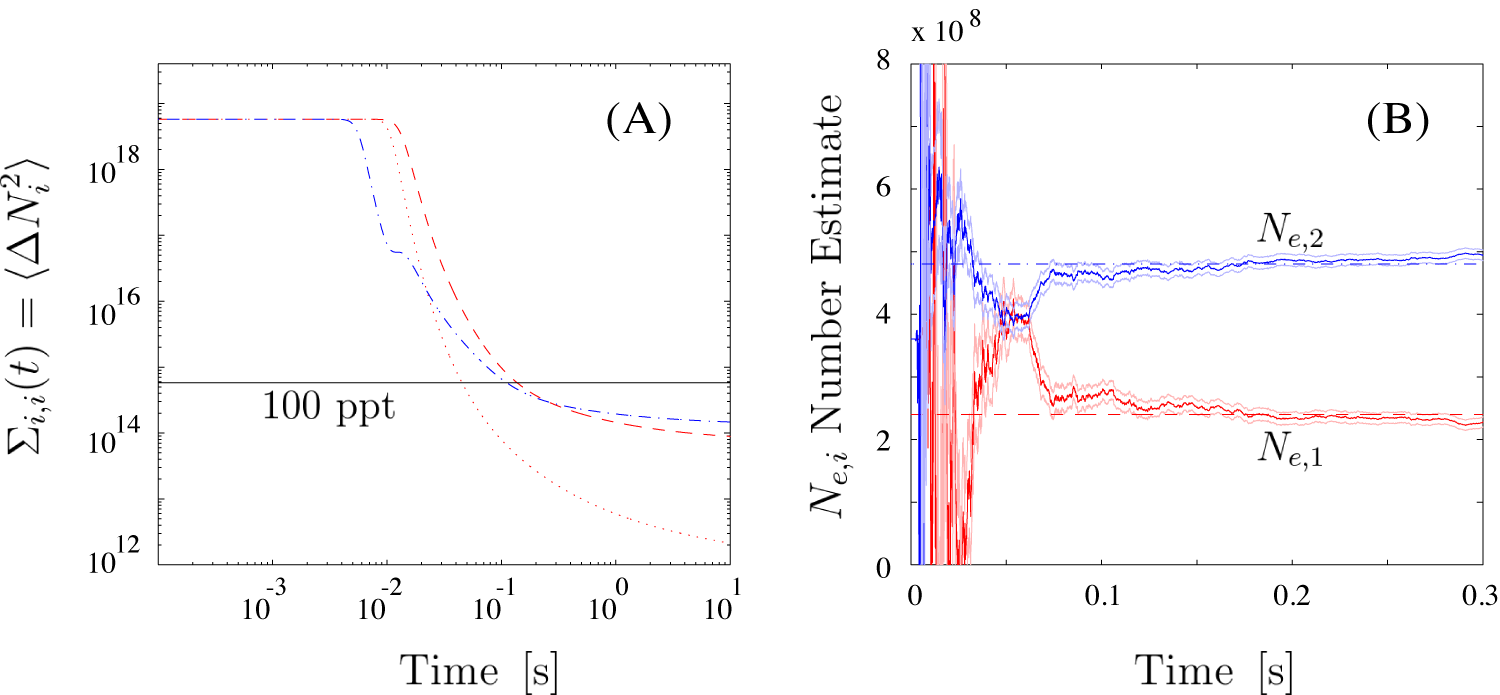}
\caption{(A) The Riccati solution of \refeqn{RiccatiSolution} for the scenario and parameters of fig. 1, but with $\sigma_M=10^{-11}\ \textrm{cm}^{-1}/\sqrt{\textrm{Hz}}$.  The number uncertainty for gas-1 is plotted as a dashed (red) line, and for gas-2 as a dashed-dotted (blue) line.  The horizontal line denotes a target $100$ ppt (parts per trillion) detection level.  If there were no uncertainty in the population of gas-2, the estimation performance for gas-1 would improve to that shown by the dotted (red) line.  If the diffusion constants were equal or the gases were immobile, the estimation performance would hardly improve upon the initial uncertainty.  (B) For a single trial, the population estimates as a function of time are shown.  The estimate for each is surrounded by curves separated by the Riccati solution uncertainty.  The true value for each gas is given by the horizontal lines (corresponding to $1$ and $2$ ppb), to which the estimates converge at long times.} \label{Figure::2}
\end{figure*}

The Bayesian estimator performance is shown in fig. 2.  Here we introduce two species into the chamber with equal optical scattering cross-sections ($\tilde{\sigma}_1=\tilde{\sigma}_2$), but with diffusion constants differing by a factor of two ($D_2=2 D_1$) and initially unknown concentrations.   Note that we have chosen a conservative value for $\sigma_M = 10^{-11}\ \textrm{cm}^{-1}/\sqrt{\textrm{Hz}}$, and thus significantly improved performance should be possible before reaching a typical photon shot noise limit. We assume prior uncertainty in the number of each gas at $10$ ppb, and as time increases, we determine the number in each species with increasing confidence.  The uncertainty variances eventually level off as the gas dissipates throughout the chamber and reaches a flat steady-state distribution. We recognize that in the absence of uncertainty in gas-2, the measurement of gas-1 improves as denoted by the dotted line.  However, as the uncertainty of gas-2 is increased, reflecting the possibility of a gas sample with trace gas-1 and large background concentration of gas-2, the performance of the gas-1 measurement remains near the level indicated by the dashed line.  This indicates that the method will work well to identify trace components despite the possibility of a large background concentration.

In conclusion, we have used Bayesian estimation to show that an array of absorption detectors can be a highly selective trace gas detector without the need for frequency tunability.  While we have presented a simplified Gaussian treatment for numerical purposes, a full Bayesian treatment with bimodal distributions and 3-D diffusion requires a simple extension of these concepts. For this diffusion selectivity approach, increased precision in the library of known quantities for each gas species (e.g., $D_i$, $\tilde{\sigma}_i$) under various environmental conditions would improve performance.  With the technical challenges of existing selectivity methods in trace gas detection including expense, portability, and bandwidth, our diffusion model with Bayesian estimation presents a practical alternative.

We thank A. and J. Kushner for their support.

\bibliography{Bib}

\end{document}